# A Survey of TCP Reno, New Reno and SACK over Mobile Ad-Hoc Network

Md Nazmul Islam Khan[1], Rashed Ahmed[2] and Md. Tariq Aziz[3]

[1]Department of Electrical & Computer Engineering, Presidency University, Bangladesh
`nazmulk@mail.presidency.edu.bd`
[2]School of Computing, Blekinge Institute of Technology, Karlskrona, Sweden
`rashed_babu2003@yahoo.com`
[3]School of Computing, Blekinge Institute of Technology, Karlskrona, Sweden
`tariq_ruet@yahoo.com`

## ABSTRACT

*Transmission Control Protocol (TCP) is often preferred to be implemented at the transport layer of a Mobile Ad-hoc Network (MANET) because of its wide range of applications, which enjoys the advantage of reliable data transmission in the Internet. However, because of some unique characteristics of MANET, TCP cannot offer reliable services while using e-mail, internet search and file transmission in such a network. The research investigates how well the different versions of TCP respond to various performance differentials when subjected to different network stresses and topology changes, aside from identifying the most efficient and robust TCP version(s) for different MANET scenarios. Among several TCP variants, three types are considered important for the analysis, namely TCP Reno, TCP New Reno and TCP Selective Acknowledgment (SACK). In most cases, the TCP performance is found in our study to decrease when the node size and mobility rate is increased in the network. There is, however, exception to this. As our simulation results demonstrate, the increases in the node velocity sometimes help the TCP to attain a better performance. The study also reveals that out of the three variants, TCP SACK can adapt relatively well to the changing network sizes while TCP Reno performs most robustly in the presence of different mobility rates within MANET.*

## 1. INTRODUCTION

A Mobile Ad-hoc Network (MANET) is a collection of mobile devices dynamically forming a communication network without any centralized control and pre-existing network infrastructure. Since their inception within the past decade, MANET has received significant attention in the world of computer research in view of their potential application in many domains. Particular applications of MANET include scenarios where infrastructure is expensive to set up and difficult or even impossible to deploy, such as battle field or rescue operations [1].

A transport layer protocol like Transmission Control Protocol (TCP) is needed in the network to establish a reliable end-to-end connection over the unreliable Internet Protocol (IP). Now-a-days, most of the internet traffic is carried out as well as the majority of widely used applications are provided by TCP [2]. Applications like File Transfer Protocol (FTP), Hypertext Transfer Protocol (HTTP) make use of TCP of TCP/IP suite for their operation. Consequently, TCP is often preferred to be implemented at the transport layer of MANET which, eventually, facilitates in connecting to the Internet, thereby providing a large extent of applications. Despite that TCP is expected to have a trust-worthy and stable performance in MANET contemporary research literature reveals that TCP does not ensure that much reliable services in such a network in applications like e-mail, internet search and file transfer, as much as it does in a wired network [3].

There are several factors that influence TCP performance in MANET, such as dynamic topology, shared medium, high Bit Error Rate (BER) and signal fading [4]. For dynamic topology, the packet losses occur due to the broken routes between the nodes whereas TCP assumes that the losses are due to the network congestion. Therefore, the network undergoes the counterproductive invocation of congestion control mechanisms employed by the TCP [5]. Additionally, the network experiences hidden and exposed node-problems due to the shared medium, thereby resulting in significant performance degradation in the network. Similarly, there are other types of constraints that have to be encountered when TCP is analyzed in the MANET environment. Therefore, a detailed analysis is required in order to gain an insight of these factors. In recent years, one of the purposes of existing research is to make improvement to the overall TCP performance for MANET scenario. However, prior to making such improvement, it is worthwhile to investigate as to what extent the TCP performance is degraded in MANET when subjected to different network stresses and topology changes. Thus, the study of TCP performance along with the investigation of the main factors affecting the TCP performance in MANET has become an important area of contemporary research. In this respect, the current study has contemplated three TCP variants in order to assess their performance over various stressful and dynamic scenarios in MANET, which eventually ascertain the relative performance merits of each TCP variant for those scenarios. The considered variants are TCP Reno, TCP New Reno and TCP Selective Acknowledgment (SACK). These three variants are reckoned as the most prominent transport layer mechanisms, which offer standard window based congestion control algorithms.

Along with the traditional difficulties of wireless environment, MANET includes further challenges to TCP. Following this, in this study, we begin by addressing the main challenges affecting the performance of TCP in a MANET environment. Subsequently, the research investigates how well the proposed three TCP variants respond to various performance differentials, such as download response time, upload response time and retransmission attempt, aside from identifying the most suitable TCP version(s) for a specific routing protocol in different network scenarios. Such analysis is important since it facilitates in determining the most suitable and robust TCP variant in a bid to optimizing the traffic goals in respective networks. A number of important system parameters such as network size and node mobility rate are taken into consideration in our study. The changes of such parameters are made (i.e. small, medium and large size network and low, medium and high node speed) to realize different realistic MANET scenarios as well as to gauge the extent of their impact on the performance of transport layer protocols.

Some attempts were previously made at describing TCP behaviour in conventional wireless networks. However, research works addressing the performance of most popular TCP variants in MANET have been lacking. More specifically, very few studies have been carried out on aspects relating to the performance evaluation of TCP variants under various node sizes and mobility conditions. To the best of our knowledge, this study is first of its kind, in undertaking experiment through analyzing the performance of three TCP variants (Reno, New Reno and SACK) in terms of upload response time, retransmission attempts, and download response time within MANET. TCP optimization in MANET has been investigated in several studies (e.g. [1], [6], and [7]). Likewise, some researches (e.g. [8], and [9]) have addressed TCP performance problems caused by the route failures in a MANET. However, from their study, it is a bit unclear as to at what extent the performance degradation of TCP occurs in MANET. In [10] Bhanumathi and Dhanasekaran conducted a simulation study of TCP Reno, Westwood and BIC-TCP, where the results clearly demonstrated the superiority of Reno variant over the other TCP versions. However, the study lacked in realizing different realistic scenarios and also a single source of TCP traffic is simulated in their study. A similar type of research is also conducted in [11] by Najiminaini, Trajković and Subedi, where the performance evaluation was made for Reno, Tahoe, New Reno and SACK versions. The research revealed that in the presence of signal attenuation, fading, and multipath, TCP Reno outperformed other congestion control algorithms in terms of congestion window, throughput and goodput. The experimental results, however, are obtained running simulation for a typical wireless network, not for a MANET. Apart from that, the study utilized low load traffics in the network. One can observe the difference in that our study explores the robustness of the TCP variant(s) by employing heavy congestions with high load traffics for both FTP and HTTP. The rest of the paper is structured as follows. Section 2 deals with identifying constraints that affect TCP performance in a MANET environment. Section 3 describes the task of MANET modelling and experimental designs. In section 4, a discussion on the results obtained upon running the simulation experiments is documented. Finally, the conclusions along with exploring avenues for future research are drawn in section 5.

## 2. TCP PERFORMANCE IN MANET

Even though TCP ensures reliable end-to-end message transmission over wired networks, a number of existing researches have showed that TCP performance can be substantially degraded in MANET. This section continues with a description of different types of constraints influencing TCP performance in MANET.

### 2.1. Route Failure

In MANET, the mobility of the node is considered as the major reason for the route failure and the route reestablishment is instantly needed in case of route failure. However, it is likely that a new route establishment may experience longer duration than the RTO of the sender. In consequence of that, the TCP sender will unnecessary invoke congestion control mechanism.

### 2.2. Path Asymmetry Impact

The network topology is changed very frequently and arbitrarily within MANET, which leads to the creation of an asymmetric path. This path formation negatively influences the TCP performance since TCP is highly dependent on time responsive feedback information. The sender starts transmitting data in a burst when a number of ACKs are received together, which causes the packet to be lost. In MANET, path asymmetry can be grouped into different forms such as loss rate asymmetry, bandwidth asymmetry and route asymmetry.

### 2.3. Network Partitioning

A network partition takes place when a node departs from the network, resulting in an isolation of some parts of a mobile ad-hoc network. These fragmented portions are defined as partitions. In MANET, TCP considers network partitioning as one of the most imperative challenges which is mainly caused due to the mobility or energy-constrained (limited battery power) operation of nodes. When the source and the destination of a TCP connection lie in different parts of the network, all transmitting packets are found to be dropped by the network. As a result, the congestion control algorithm will be invoked instantly by the TCP sender.

### 2.4. Hidden and Exposed Node Impact

Figure 1 presents a typical hidden node condition where packet transmission starts from node A to node E. Since, node B cannot sense node D, node B assumes the channel as an idle channel and therefore initiates its transmission by dispatching a Request to Send (RTS) to node C. However, transmitting RTS unexpectedly introduces collisions because node C is found in the interference range of node D. This problem is termed as "Hidden Node" impact where node D is called the hidden node with respect to node B.

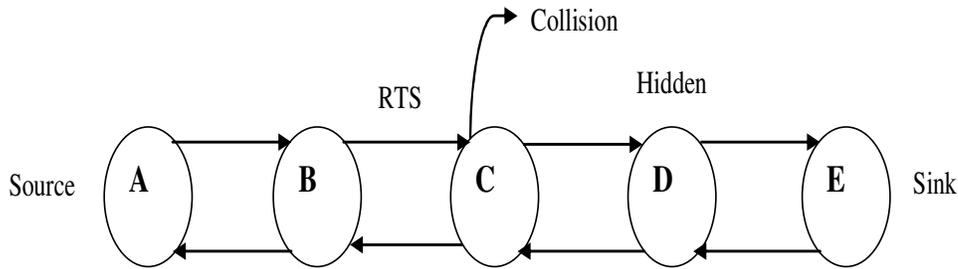

Figure 1. Hidden node impact

Figure 2 depicts a condition through which the exposed node problem can be realized. When node D intends to transmitting data toward node E, node C will not be able to send any data frame to node B. Node C must wait until node D finishes its current transmission to node E. This is because node D is within the sensing range of node C. This problem is known as "Exposed Node" impact where node D is called the exposed node with respect to node C.

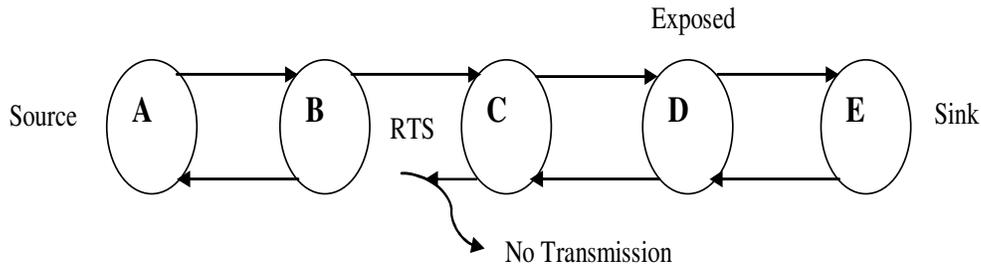

Figure 2. Exposed Node Impact

## 3. EXPERIMENTAL DESIGN

This section comprises a description of the designed network model and the necessary parameters, which are realized in configuring the network model.

### 3.1. Evaluation Platform

The research is conducted using discrete event simulation software known as Optimized Network Evaluation Tool (OPNET) [12], which is just one of several tools provided from the OPNET Technologies suite. In order to undertake the experimental evaluation, the most recently available version, namely OPNET Modeller 16 has been adopted in our study.

### 3.2. Performance Metrics

To evaluate the performance of the TCP variants, upload response time, download response time and retransmission attempts are considered as performance metrics. The efficiency and effectiveness of upload and download activities are evaluated by the extent of upload response time and download response time, respectively. Hence, in data traffic measurements these two quantified parameters play a vital role where the lower the value achieved, the faster the task proceeded. Retransmission attempts, on the other hand, can be defined as the total number of retransmission attempts of packets that have been lost or damaged due to a link failure in the network. It also shows the number of packets failed in the process, which, in effect, requires retransmission. So, the lower is the retransmission attempts the more reliable is the TCP variant.

## 3.3. Network Modelling

The network models of the current study are designed, in the OPNET simulator, by taking help of different network entities. The network entities used during the design of the network model are wireless server, application configuration, profile configuration, mobility configuration and workstations (nodes). These model objects are basically a series of network components that allow attribute definition and tuning.

Application configuration is an essential object that defines the transmitted data, file size and traffic load. More often, it supports common applications, namely, HTTP, FTP, Database, Email, Print and so on. We have chosen FTP and HTTP applications for data traffic analysis where each application is considered with heavy traffic load (individually), in line with the requirement for bandwidth utilization. On the other hand, profile configuration is employed to create the user profiles whereas these profiles are specified on different nodes for generating the application traffic. For instance, an FTP profile is created in a profile configuration entity in order to support the FTP traffic, which is generated by an application configuration entity.

One of the other important entities is the mobility configuration, which is used for the purpose of determining the mobility model of the nodes. Moreover, it has to select several appropriate parameters such as speed start time, stop time, pause time and the like, to properly control the movement of the nodes in the network. The reason for configuring the mobility object is to allow the nodes to move within the specific allocated network area, which is chosen as 1000 square meters in our study. In other words, the traffic generated from outside this specific range, if any, will not be taken into account. Meanwhile, to configure the nodes with a mobility option, a widely used mobility model known as the default random waypoint mobility is used for all simulation purposes in the present study. The combination of pause time and velocity sets up relative degrees of mobility between mobile nodes in the simulated network. To symbolize the mobile behaviour of the nodes, the speed of the node is initially set to 10 m/s with a pause time of 50 sec to observe the network behaviour with low mobility. At some later stage, the speed is increased to 20 and 30 m/s with the same pause time so that the nodes can travel with greater speed in the network. The reason for increasing the node speed is to observe the impact of mobility on the network performance. The server module is basically a WLAN server, which is configured to support and control the application services (i.e. FTP and HTTP) based on the user profile. The connection speed is set at 5.5 Mbps in our study. Finally, all mobile nodes are configured to generate FTP and HTTP traffic randomly, with the ability to route the data packets to the desired destinations.

Table 1 demonstrates the general parameters used in the process of all simulation experiments of the study. Meanwhile, the parameters used for wireless LAN configuration are portrayed in Table 2. The buffer size is set to 256,000 bits because a medium flow of application has been intended to be generated in our experiment. Likewise, in order to avoid the potential problem related to manual error, the channel setting is fixed at that which is auto-assigned. The channel setting parameter is important since it specifies the bandwidth of the radio channel for physical layer transmissions. This auto-assigned option sets the bandwidth to 22 MHz. On the other hand, the parameters such as slow start initial count, initial RTO, minimum RTO and maximum RTO are used as TCP simulation parameters, and their values are shown in Table 3. Meanwhile, as explained in the OPNET product specifications, the start time for a file transfer session is computed by adding the inter request time to the time that the previous file transfer started. Following that, we specified the profile generation start time as uniform (100, 110) and the start time for the application is set as constant (5), as shown in Table 4 and Table 5, respectively.

Table 1. General parameters.

| General parameters | Value |
| --- | --- |
| Area | 1000x1000 square meters |
| Network size | 30, 60 and 100 nodes |
| Data rate | 5.5 Mbps |
| Mobility model | Random way point |
| File size | High load |
| Traffic type | FTP, HTTP |
| Mobility speed | 10, 20 and 30 m/s |
| Simulation time | 600 seconds |
| Address mode | IPv4 |

Table 2. Wireless LAN parameters.

| Wireless LAN parameters | Value |
| --- | --- |
| Wireless LAN MAC address | Auto assigned |
| BSS identifier | Auto assigned |
| Physical characteristics | Direct sequence |
| Data rate (bps) | 5.5 Mbps |
| Channel settings | Auto assigned |
| Transmit power(W) | 0.005 |
| Rts threshold (bytes) | None |
| Fragmentation threshold (bytes) | 1024 |
| CTS-to-self option | Enabled |
| Short retry limit | 7 |
| Long retry limit | 4 |
| AP beacon interval (seconds) | 0.02 |
| Max receive lifetime (seconds) | 0.5 |
| Buffer size (bits) | 256000 |
| Large packet processing | Fragment |
| PCF parameters | Disabled |

International Journal of Distributed and Parallel Systems (IJDPS) Vol.3, No.1, January 2012Table 3. TCP parameters.

| TCP parameters | Value |
|---|---|
| Slow start initial count | 1 |
| Receive buffer size (bytes) | 8,760 |
| Maximum ACK segment | 2 |
| Duplicate ACK threshold | 3 |
| Initial RTO (seconds) | 1.0 |
| Minimum RTO (seconds) | 0.5 |
| Maximum RTO (seconds) | 64 |
| RTT gain | 0.125 |
| Deviation gain | 0.25 |

Table 4. Profile configuration parameters.

| Profile configuration | Value |
|---|---|
| Number of profile | 2 (FTP and HTTP) |
| Operation mode | Simultaneous |
| Start time (seconds) | Uniform (100,110) |
| Duration (seconds) | End of simulation |
| Profile repeatability | Once at start time |
| Inter-repetition time (seconds) | Constant (300) |
| Number of repetitions | Constant (0) |
| Repetition pattern | Serial |

Table 5. Application configuration parameters.

| Application configuration | Value |
|---|---|
| Number of application | 2 (FTP and HTTP) |
| Start time offset (seconds) | Constant (5) |
| Duration (seconds) | End of profile |
| Application repeatability | Once at start time |
| Inter-repetition time (seconds) | Constant (300) |
| Number of repetitions | Constant (0) |
| Repetition pattern | Serial |

55

## 4. RESULT AND ANALYSIS

This section presents details of the experiments carried out to evaluating TCP performance when the node sizes and mobility rates are varied in a MANET environment. We considered total simulation time as 600 seconds over which the performance statistics of upload response time, retransmission attempts and download response time are collected. In order to achieve the simulation accuracy in OPNET, five replications are run for each experiment, with different constant seeds of the Pseudo Random Number Generator (PRNG). The constant values of the seeds are used since it minimizes the variance of the simulation results and thus allows a better comparison of the TCP variants. In OPNET, the PRNG is supported by Berkeley Software Distribution (BSD)'s algorithm, providing safe random numbers. We understand that five simulation replications were reasonably sufficient since all such replications portray quite similar graphical results. Thus, the interpretation and analysis of the results led us to the same conclusion.

### 4.1. Varying Node Size

To observe the impact of node variation on the performance of TCP Reno, TCP New Reno and TCP SACK, the target applications are run with various network sizes (30, 60 and 100 nodes). Dynamic Source Routing (DSR) protocol is used as a network layer protocol since DSR interacts with TCP more efficiently than the other routing protocols in MANET, as it is stated in [7]. Though this section deals with network size issue, it is much more realistic for a MANET environment to generate at least a low mobility rate instead of keeping it fully static. Accordingly, a moving speed of 10 m/s with an average pause time of 100 sec is set to allow the mobile nodes to move slowly in the network.

Figure 3 demonstrates the download response time of three TCP versions for transmitting an FTP file (50,000 bytes). In a small network (e.g. 30 nodes), all the three versions achieve almost an equal response time, however, with a slightly better performance in case of Reno. In a medium network (e.g. 60 nodes), Reno continues achieving the shortest response time in acquiring the FTP file. However, in a large network (e.g. 100 nodes), SACK reduces its response time dramatically and outperforms other TCP variants. It is apparent that the performance of all the TCP versions degrades when more nodes are added in the network. In the presence of large number of users, the network becomes more congested. Other than the congestion problem, the network is likely to become more prone to signal attenuation and channel error due to the establishment of more links among the users. This causes TCP to unnecessarily invoke the counterproductive and time consuming congestion control mechanisms. And due to such invocation, more time is taken to finish the data recovery activities and hence more time is to be spent to download a file in the presence of high number of nodes.

The download response time is highly influenced by TCP's congestion window size. More especially, the larger the congestion window size, the shorter would be the file response time. Meanwhile, it was investigated from [11] that in the presence of heavy network congestion and high signal attenuation, TCP SACK and Reno often maintain a larger congestion window size, compared to that in other TCP variants. Thus, a validated simulation is accomplished, since our study demonstrates that both TCP SACK and Reno achieve a shorter file response time in a high density network.

Figure 4 shows the performance of three TCP versions in terms of upload response time. In a small (e.g. 30 nodes) and a medium (e.g. 60 nodes) network, New Reno ensures the lowest upload response time, approximately 0.19 and 0.46 sec, respectively. Meanwhile, the SACK outperforms other TCP versions in a large network (e.g. 100 nodes) by taking the lowest response time of about 1.12 sec. It appears that the small and medium networks accommodate the FTP file relatively quickly and thus ensure a smooth transmission. In such networks, the average response time takes less than one second for all the variants. However, due to uplink limitation a much higher response is required in a situation where a large number of users are present (e.g. 100 nodes). When all 100 nodes start uploading concurrently, extra load is gained from the large network. Apart from that, due to the presence of a high packet error rate in a large MANET, the TCP retransmission mechanism is generated more frequently, thereby consuming more network bandwidth. This eventually leads to huge delay to upload an FTP file in a high density network. The difference between download and upload response time is not always obvious. However, the response time differs in our simulation results. This is due to the fact that the bandwidth is asymmetrical and the available throughput is greatly limited in MANET. In a network where asymmetric connection is supported, the rated speeds of upload usually differ from that of download. Hence, the upload and download response time might not be same. Download activities generally can be performed faster than that of upload activities. However, as our results demonstrate, the opposite result is also possible. In our study, the upload response time of an FTP (50000 bytes) file is always shorter than that of the download response time. When the connection experiences a heavy user load, the download response time might suffer. Note that we observed the majority of the nodes in the network were engaged in the downloading of files instead of in the uploading.

Figure 5 shows the performance of three TCP variants in terms of their retransmission attempts. With the increase in node numbers, the numbers of retransmission attempts are increased for all the three versions. This is due to the physical layer disconnections as well as the increase in packet error rates in a high density network. Likewise, the channel contention is also increased as more routing loads are experienced in larger networks, thereby resulting in more packet losses. In response to a packet loss in a traditional network, TCP retransmits the lost packet from its own source. But, in a MANET associated with a high error rate, TCP may have to take several retransmissions to deliver a single packet to its destination successfully. Among the three variants, TCP Reno dominates in both small and medium density networks, maintaining the lowest average retransmission rate of 0.024 packets/sec and 0.05 packets/sec, respectively. Meanwhile, the performance of SACK is found to be quite impressive in a large network as it attains a relatively lower retransmission rate (0.085 packets/sec), compared to those of the other variants. On the other hand, one can notice that the performance of New Reno degrades significantly in a large size network. In such a network, aggressive employment of window mechanisms is considered one of the main factors for causing more retransmissions by New Reno. During the slow start phase, the aggressive and inappropriate window growth of New Reno causes the network to be overloaded, which induces periodic packet losses on the link layer and more frequent timeouts in the transport layer. Thus, frequent link contentions and more link failures are occurred in the MAC layers and resulted in a great number of retransmission in New Reno based network.

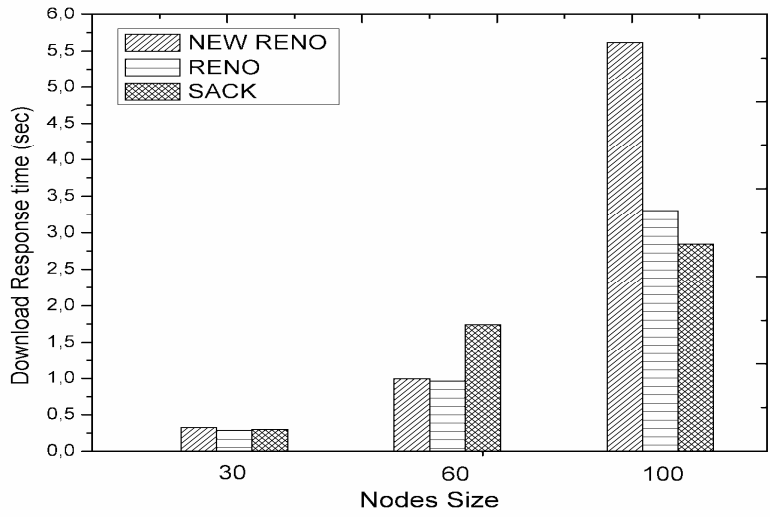

Figure 3. Average download response time for different sizes of nodes

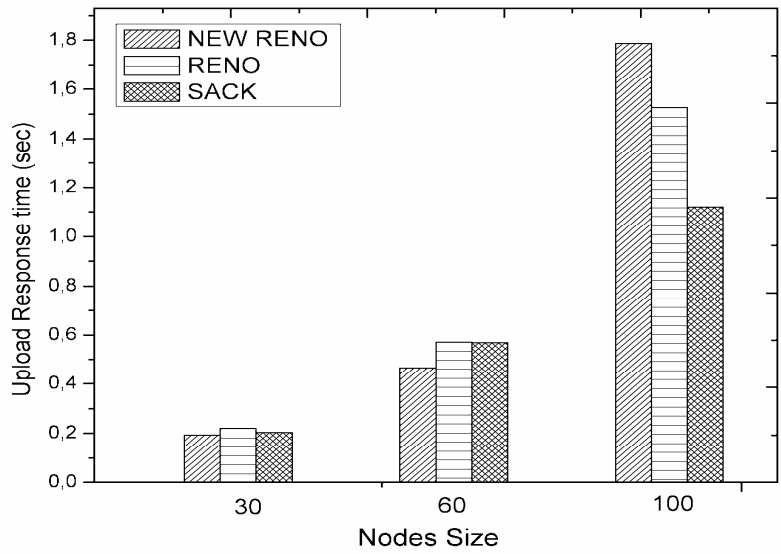

Figure 4. Average upload response time for different sizes of nodes

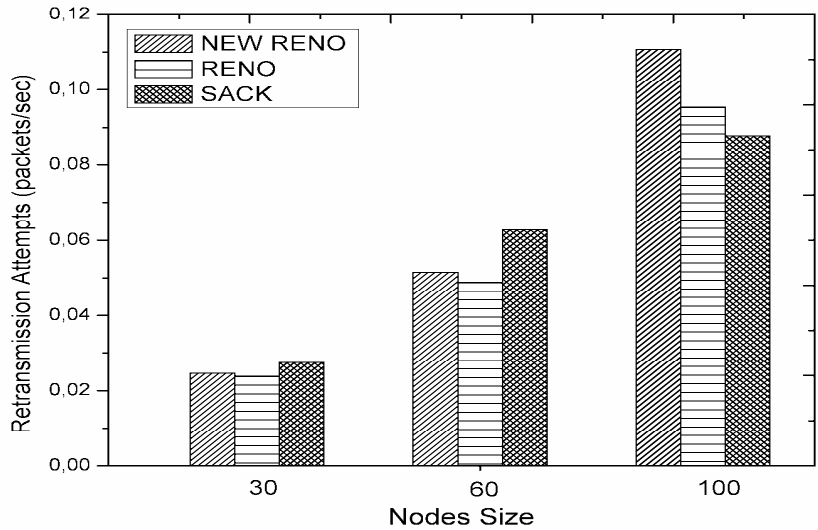

Figure 5. Average retransmission attempt for different sizes of nodes

## 4.2. Varying Node Speed

This section presents the performance results of TCP variants obtained upon running the experiments with various mobility rates. The scenarios considered in the analysis consist of 60 nodes moving with node speeds of 10, 20 and 30 m/s. The pause time is set to 50 sec for all node speeds. The purpose of such experiments is to evaluate TCP performance when the rate of topology changes in MANET.

In Figure 6, the performance of different TCP versions is analyzed in accordance with the download response time. In all the scenarios (10, 20 and 30 m/s), Reno variant achieves the lowest download response time compared to others.

With the incidence of increased mobility rates, frequent changes of the nodes occur, subsequently causing frequent changes in the link states and further more packet losses. When link failure takes place owing to the mobility, all of three TCP versions mostly differentiate the packet loss through observing the TCP RTO timer. Since none of the versions are designed to cope with such situations (i.e. link failures), they all are supposed to react similarly in a MANET. In fact, the TCP performance is expected to be dramatically affected. Nevertheless, in our study, the changes in mobility rate to higher values result in a variety of reactions to three TCP versions. When the node speed is changed from 10 to 20 m/s, the average download response time of all the TCP variants, except for SACK, is found to increase. Surprisingly though, the response time of all the TCP versions decreases when the speed is changed to 30 m/s. It appears that the motion helps the network to discover routes and to achieve a better connectivity. This leads one to conclude that the increasing node velocity does not always work as a degradation factor for the TCP performance in a wireless environment. It is true that increasing the node speed to 30 m/s increases the probability of frequent changes in the network topology and frequent link breakages. However, at the same time, it makes it possible for the ad-hoc routing protocol (e.g. DSR in our case) to re-establish the link faster than the RTO duration. When the time required for re-establishing a broken link is shorter than the RTO, the TCP experiences no packet loss and consequently it does not trigger the time-consuming congestion control mechanisms. This eventually leads the TCP to exhibit a better performance in the network. However, it is not obvious that furthering the node velocity will keep reducing the response time. Instead, it might increase the response time to a higher extent. Therefore, the choice of a right mobility rate within MANET can be considered as an important area of research.

The results of upload response time against different mobility rates are analyzed in Figure 7. In a 10 m/s network, the lowest response time is observed in Reno version, amounting to approximately 0.47 sec. For a 20 m/s network, Reno and New Reno require shorter response time, approximately 0.51 and 0.52 sec, respectively. However, when the node speed is changed to 30 m/s, New Reno is no longer able to perform superiorly; the performance is rather deteriorated abruptly. Reno version, on the other hand, continues to dominate in a high mobility network by ensuring the lowest response time of about 0.44 sec. Meanwhile, in the case of SACK variant, a moderate response time is achieved in all the scenarios, although the performance tends to be outperforming in the long run.

Finally, Figure 8 demonstrates the retransmission attempts in the presence of different mobility rates within MANET. As one can observe, for all three TCP versions, the increase in the node speeds up to 20 m/s results in a decrease in retransmissions, whereas the increase in node speeds up to 30 m/s results in an increase in retransmissions. The Reno version shows an outstanding performance in 10 and 20 m/s scenarios through achieving the lowest retransmission rate among the three. However, when the node mobility is shifted to 30 m/s, slightly higher retransmissions are attempted by Reno, amounting to approximately 0.051 packets/sec. In such a network, the highest retransmission is attempted with the New Reno variant which is about 0.067 packets /sec. On the other hand, the SACK version accounts for the lowest retransmission in such a network, approximately 0.046 packets/sec, which is about 1.07 and 1.46 times as less as that of Reno and New Reno variants, respectively. Looking at the scenarios, it is evident that SACK variant is relatively more robust to the dynamics of the wireless channels. This version allows the receiver to identify only the segments that are received, hence the sender usually retransmits only the lost segments, resulting in a less number of retransmission attempts compared to in the other two versions.

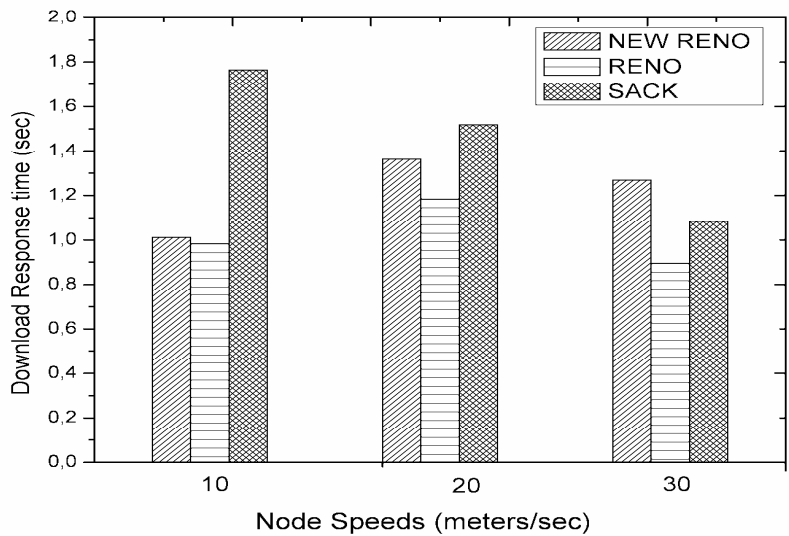

Figure 6. Average download response time for different node speeds

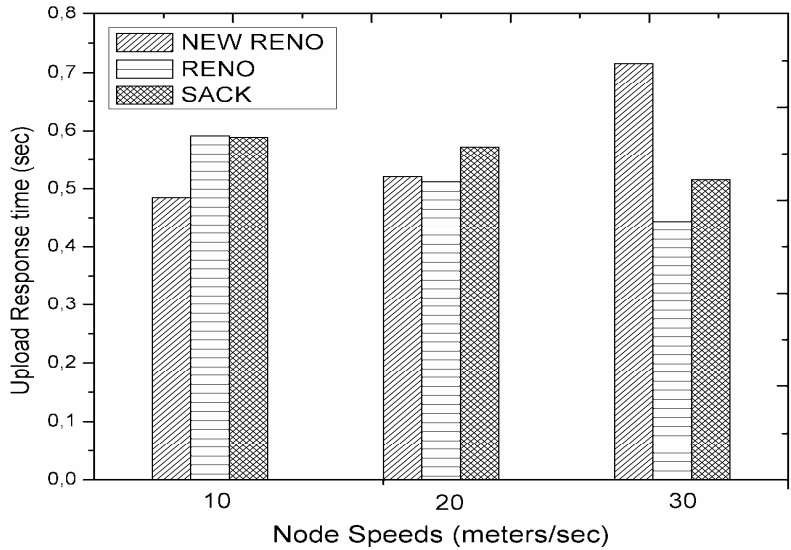

Figure 7. Average upload response time for different node speeds

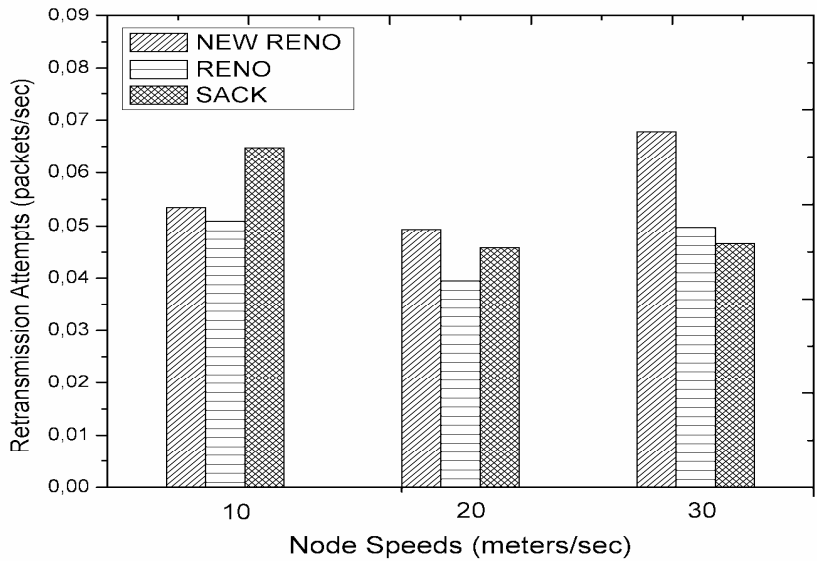

Figure 8. Average retransmission attempts for different node speeds

## 5. CONCLUSIONS

This research makes contribution in three areas. Firstly, the research undertakes an analysis of different types of constraints influencing TCP performance in MANET. Secondly, the study analyzes the performance of the three most widely used TCP variants (Reno, New Reno and SACK) in an ad-hoc environment. In this respect, an investigation is made into aspects as to how well these variants respond to different network conditions, particularly with respect to extension of network size and variation of mobility rate. Finally, using the simulation environment, an analysis is carried out on the results of upload response time, download response time and retransmission attempts. These results have facilitated in determining the most suitable TCP variants that can perform more efficiently and robustly in various MANET scenarios. The key observations of the research are as follows.

In a high density network when congestions and packet error rates are very likely, TCP SACK outperforms other variants in terms of retransmission attempts, upload and download response time. The performance of Reno is also noteworthy, which is, however, limited to a small and a medium size network. Meanwhile, with the variations of mobility rates, TCP Reno dominates other congestion control algorithms in most of the cases. However, the performance of SACK is also remarkable in a high mobility scenario. Particularly in terms of retransmission attempts, the SACK variant demonstrates its superiority over the other versions. On the other hand, New Reno is found to be a less suitable variant under higher network stresses and mobility conditions.

The study also reveals some interesting findings on TCP variants when their performances are evaluated over dynamic topologies in MANET. It has been observed that the performance of all TCP versions studied in this research decreases when the number of nodes is increased in the network. Likewise, in most cases, the TCP performance is found to decrease when the mobility rate is increased in the network. There is, however, exception to this. It has been noticed that the increase in the node velocity sometimes help the TCP to attain a better performance. Essentially, the performance results among three TCP variants are found to be highly fluctuating with the increase in mobility. The performance varies from speed to speed, sometimes following an increasing trend while sometimes following a decreasing trend. In some cases, it was a bit unclear as to how the curves might behave for further increases in the node speeds. Hence, in any future study, the determination of an ideal mobility rate for TCP variants within MANET is worth pursuing. In our study, we have considered two network factors (node size and mobility); the pursuit of future research may include aspects relating to evaluation of TCP performance under other important factors like network load and transmission range.

## REFERENCES

[1]     Saher S. Manaseer, "On backoff mechanisms for wireless mobile ad hoc networks," in Ph. D thesis, The Faculty of Information and Mathematical Sciences at University of Glasgow, Scotland, 2009, PP. 1-156.

[2]     D. Triantafyllidou and K. Al Agha, "Evaluation of TCP performance in MANETs using an optimized scalable simulation model," in 15th International Symposium on Modeling, Analysis, and Simulation of Computer and Telecommunication Systems, 2007, MASCOTS '07, pp. 31-37, November 2008.

[3]     I. Chlamtac, M. Conti and J. Liu, "Mobile ad hoc networking: imperatives and challenges," Ad Hoc Networks Journal, vol.1, no. 1, pp. 13-64, Jul. 2003.

[4]     O. Bazan, U. Qureshi, M. Jaseemuddin and H.M. El-Sayed, "Performance Evaluation of TCP in mobile ad-hoc networks," in The Second International Conference on Innovations in Information Technology, IIT'05, Toronto, Canada, 2005, pp. 175-185.

[5]     Berger, Lima, Manoussakis, Pulgarin and Sanchez, "A performance comparison of TCP protocols over mobile wireless ad hoc networks," IEEE Electronics, Robotics and Automotive Mechanics Conference, Vol.2 pp.88-94, 2006.

[6]     S. B. Lee, G. S. Ahn, and A.T. Campbell, "Improving UDP and TCP performance in mobile ad hoc networks with INSIGNIA," IEEE Communications Magazine, vol. 39, no. 6, pp. 156-165, August 2002.

[7]     K.Kathiravan, S. Thamarai Selvi, and A.Selvam, "TCP performance analysis for mobile adhoc network using on-demand routing protocols," Ubiquitous Computing and Communication Journal, pp. 370-376, April 2007.

[8]     S. Papanastasiou, M. Ould-Khaoua, and L. Mackenzie, "On the evaluation of TCP in MANETs," in Proceedings of the International Workshop on Wireless Ad Hoc, August 2005.

[9]     A. Al Hanbali, E. Altman and P. Nain, "A survey of TCP over mobile ad hoc networks," Research Report no. 5182, INRIA Sophia Antipolis research unit, May 2004.

[10]    V. Bhanumathi and R. Dhanasekaran, "TCP variants - A comparative analysis for high bandwidth - delay product in mobile adhoc network," in 2nd International Conference on Computer and Automation Engineering (ICCAE),2010, Singapore, 2010, pp. 600-604.

[11]    Mohamadreza Najiminaini, Ljiljana Trajković and Laxmi Subedi, "Performance evaluation of TCP Tahoe, Reno, Reno with SACK, and NewReno using OPNET modeler," in  Simon Fraser University Vancouver, British Columbia, Canada, 2009.

[12]    OPNET Simulator, Retrieved 15 June, 2010, [Online], Available:  http://www.opnet.com

**Authors:**

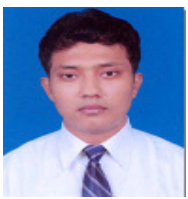
**Md Nazmul Islam Khan** was born in Dhaka, Bangladesh in 1985. He obtained his M.Sc degree in electrical engineering from Blekinge Institute of Technology, Sweden in 2010. Currently he is working as a faculty at Presidency University, Bangladesh. His research interest includes Wireless communication, OFDM and multicarrier systems, radio propagation, and MANET.

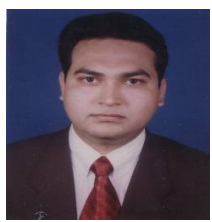
**Rashed Ahmed** was born in Pabna, Bangladesh in 1983. He obtained his B.Sc. Degree in computer science from International Islamic University Chittagong (IIUC), Bangladesh .He worked as a lecturer in Ideal College, Bangladesh from November, 2005 to February, 2007. He has also experiences working in (as a system engineer) installation, commissioning and maintenance of telecommunication field at Ring Tech ltd, Bangladesh. He completed his MSc in Electrical Engineering in 2010 with emphasis on Telecommunication from Blekinge Institute of Technology (BTH), Sweden. His research interests include wireless networks, radio communication, LTE, routing protocols and MANET.

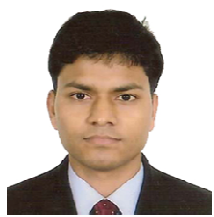
**Md. Tariq Aziz** was born in Chapai Nawabgonj, Rajshahi, Bangladesh in 1983. He completed his Bachelor Degree in Computer Science & Engineering from Rajshahi University of Engineering and Technology (RUET), Rajshahi, Bangladesh in September, 2005. He started his career as an interne in Mantrust Multimedia System Ltd. Later he worked as a support engineer in Global Online Services Ltd. (ISP) for six months. In June, 2006, he joined Signal Mountain Networks BD Ltd. (SMN), where he served as a Network Support Engineer in their Network Operation Centre (NOC). During his professional career, he realized the best way of visualizing and progressing professionally in the field of state-of-art technology would be doing further higher studies and become involved in the research field. This recognition fuelled his ambition to pursue an MSc at Blekinge Institute of Engineering (BIT), Sweden in August, 2008. He achieved his Masters degree in Electrical Engineering with emphasis on Internet Systems from Blekinge Institute of Technology (BIT), in August of 2011. His research interest includes wireless networks, LTE, QoS, MPLS, routing protocols and MANET.